\documentclass[aps,prl,reprint,superscriptaddress]{revtex4-2}
\usepackage[utf8]{inputenc}

\usepackage[colorlinks=true,citecolor=blue,linkcolor=black]{hyperref}
\usepackage{amsmath}
\usepackage{amssymb}
\usepackage{bm}
\usepackage{graphicx}  

\newcommand{\tr}{\operatorname{tr}}
\newcommand{\E}{\mathbb{E}}
\newcommand{\U}{\mathcal{U}}
\newcommand{\I}{\mathbb{I}}
\newcommand{\M}{\mathcal{M}}

\begin{document}
\title{Quantum non-Markovian noise in randomized benchmarking of spin-boson models}
\date{\today}

\author{Srilekha Gandhari}
\email[email: ]{gandhari@umd.edu}

\affiliation{Joint Center for Quantum Information and Computer Science, NIST/University of Maryland, College Park, Maryland 20742, USA}

\author{Michael J. Gullans}
\email[]{mgullans@umd.edu}
\affiliation{Joint Center for Quantum Information and Computer Science, NIST/University of Maryland, College Park, Maryland 20742, USA}

\begin{abstract}
    In non-Markovian systems, the current state of the system depends on the full or partial history of its past evolution. Owing to these time correlations, non-Markovian noise violates common assumptions in gate characterization protocols such as randomized benchmarking and gate-set tomography.  Here, we perform a case study of the effects of a quantum non-Markovian bath on qubit randomized benchmarking experiments. We consider a model consisting of qubits coupled to a multimode Bosonic bath.  We apply unitary operations on the qubits, interspersed with brief interactions with the environment governed by a Hamiltonian. Allowing for non-Markovianity in the interactions leads to clear differences in the randomized benchmarking decay curves in this model, which we analyze in detail. The Markovian model’s decay is exponential as expected whereas the model with non-Markovian interactions displays a much slower, almost polynomial, decay. We develop efficient numerical methods for this problem that we use to support our theoretical results.  These results inform efforts on incorporating quantum non-Markovian noise in the characterization and benchmarking of quantum devices.
\end{abstract}

\maketitle

Noise characterization is central to the operation of quantum devices and processors. The large dimension of Hilbert space of many-body quantum systems, coupled with the sensitivity of quantum states to measurements, makes benchmarking quantum devices a complex task. There are several methods to characterize quantum operations, with varying degrees of detail, to assess whether they’re functioning as intended. Randomized Benchmarking (RB) is a particularly common technique due its  resource-efficiency and insensitivity to state preparation and measurement (SPAM) errors. Originally created to extract average fidelity of a group of gates when the noise has no gate-dependence, it has been shown to generalize to a wide variety of other scenarios \cite{Helsen2019a, Helsen2019b, CarignanDugas2015, Brown2018, Hashagen2018, Helsen2022, Claes2021}. 

A key assumption made to simplify RB studies, as well as most other benchmarking techniques, is that of time-independent or \textit{Markovian} noise. Markovianity is a property of a stochastic process where the future of a system is dependent only on its current state irrespective of how it reached there. In other words, a Markovian quantum system does not carry any temporal correlations with information about its past. Recent studies have shown the limitations of this assumption and the need to study correlated noise \cite{Young2020, Wei2024, Aharonov2006, Kam2024, White2020,Gullans24}. Although the effects of classical correlations have been studied \cite{Groszkowski_2023,Seif2022,Huang23}, they do not capture the effects of having a quantum bath capable of entangling with the system, which is a setting gaining increased emphasis \cite{Berk2021, Berk2023, Sabale2024, deVega2017, FigueroaRomero2021}. A common approach to mitigate these effects includes dynamical decoupling \cite{Viola1998, Viola1999, Viola2003}, while the overall dynamics can be captured using several related frameworks such as quantum combs and process tensors \cite{White2020,White2022}. However, these frameworks provide a very general way to deal with non-Markovianity and give little idea about the dynamics until fully equipped with details. In this paper, we test the effects of non-Markovian noise in RB using an exactly solvable model.

Here, we consider randomized benchmarking of qudit systems coupled to a non-Markovian Bosonic bath. We provide an efficient algorithm for calculating the final state of the system after a given time. We then show that quantum non-Markovian effects have a dramatic effect on the fidelity decay by changing it from a simple exponential to a power-law times exponential. In our model the system and the bath interact through a fixed spin-Boson Hamiltonian \cite{Loss1998, Bluhm2010,Kamar2024} and correlations are generated by the virtue of the bath having dynamics at a slower time scale than the system. We find that there are strong qualitative differences in the RB decay when comparing Markovian and non-Markovian limits of the model.  We trace the origin of the non-Markovian dynamics to heating dynamics of the bath arising from coupling to the driven system.  In addition, we identify a simple fingerprint for the presence of non-Markovian noise in RB based on the distinguishability between two orthogonal input states to the same RB circuit.  These results serve as important guides in the quest to characterize and control non-Markovian noise in quantum devices.

\begin{figure}[tb]
    \includegraphics[width = .49 \textwidth]{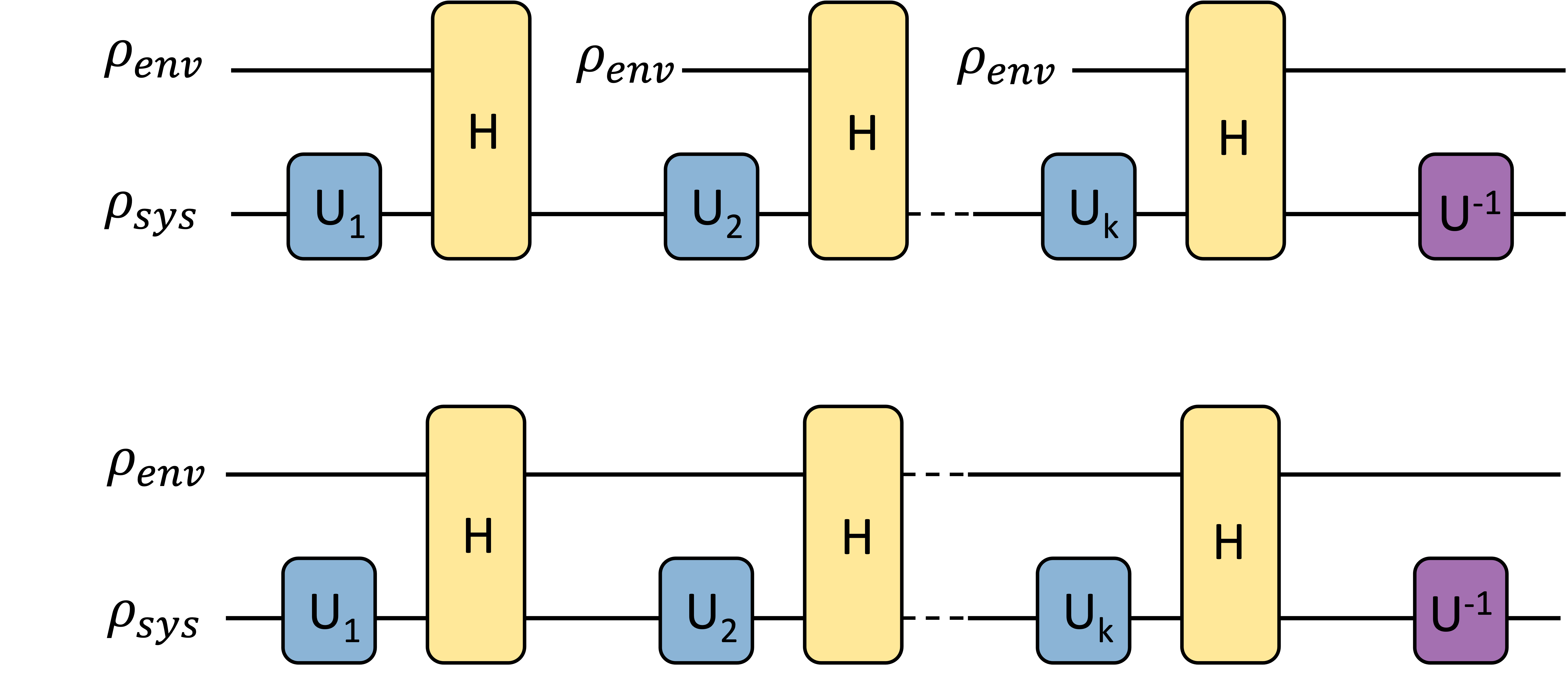}
    \caption{Circuit representation of (a) Markovian and (b) Non-Markovian dynamics in our model. All the unitaries are drawn from a two-design and $U^{-1}=(U_k\ldots U_1)^{-1}$.}
    \label{fig:circuit_model}
\end{figure}

\paragraph{Randomized Benchmarking (RB)}
RB is a common benchmarking technique that is resource efficient and resistant to state preparation and measurement (SPAM) errors. To formulate the protocol, first consider a set of gates, our \textit{gateset}, $\mathcal{G}$ that generates the Clifford group. Using RB, we can extract the average error of the gateset assuming each noisy gate can be written as a noise channel applied to an ideal gate, and that the noise is time and gate independent \cite{Emerson2005,Knill2008}. 
As shown in Fig.~\ref{fig:circuit_model}, we consider a circuit model where the qudit system $\rho_s$  interacts with a multimode Bosonic bath ($\rho_{env}$) through a Hamiltonian $\mathcal{H}$, interspersed with the unitary operations $U_1,\ldots,U_k$. In the Markovian setting, the environment is constantly refreshed in before every interaction with the system, whereas in the more general non-Markovian case, the environment continues to evolve without losing any information. The difference between the average overlap $\langle \rho_s|\rho_{final}\rangle$ of the two cases tells us the extent of non-Markovianity, that is, how much information is flowing back from the environment to affect the future of the system.

The system-environment Hamiltonian is
\begin{equation}
\begin{aligned}
      H &= \sum_{i=1}^m g_i \sigma_z x_i +\frac{\omega_i}{2} \left(x_i^2+p_i^2\right)
      \\ &=  P_0 \otimes H_0 + P_1 \otimes H_1 = H_0 \oplus H_1,
\end{aligned}
\end{equation}
where the summation is over the different Bosonic modes, $P_{0,1}$ are the projectors onto $|0\rangle$, $|1\rangle$ states, and
\begin{equation}
    H_{0/1} = \sum_i \frac{\omega_i}{2}\left(\left(x_i\pm\frac{g_i}{\omega_i}\right)^2+p_i^2\right) - \frac{g_i^2}{2\omega_i}.
\end{equation}
The evolution operator for a time duration $t$ can therefore be written as
\begin{equation}\label{eq:projectors}
\begin{aligned}
    \mathcal{H} \equiv e^{-iHt} = P_0 \otimes e^{-iH_0t} + P_1 \otimes e^{-iH_1t}~.
\end{aligned}
\end{equation}
We note that this Hamiltonian omits parts corresponding only to the system's evolution, as they can be absorbed into the applied unitary gates in the circuit. We also note that, the choice of the $Z$-basis for the coupling term is arbitrary in this model.

\paragraph{Non-Markovian Evolution}
Our objective is to find the final state of the system, after $k$ time steps, averaged over the unitary circuits. If the initial state assumed to be in a product state $\rho_{in} = \rho_s \otimes \rho_{env}$, as shown in Fig.~\ref{fig:circuit_model}(a), the final state is
\begin{equation}\label{eq:general_state}
\begin{aligned}
    \E_U\left[\rho_{k}^{sys}\right] = \mathrm{tr}_E ( \E_U[ \, \U^{-1} \circ &\mathcal{E}_H \circ \U_k \ldots 
    \\ &\circ \mathcal{E}_H \circ \U_1\, (\rho_s \otimes \rho_{env}) ])
\end{aligned}
\end{equation}
Each of the gates $\{U_i\}_{i=1}^k$ is sampled uniformly randomly from the Clifford group, with $\E_U$ denoting the expectation value over all $U_i$, and the evolution map is denoted by $\mathcal{E}_H (\cdot)= \mathcal{H}\, \cdot\, \mathcal{H}^\dagger$.

Using Eq.~\ref{eq:projectors}, with the help of strings $\textbf{p}, \textbf{q}$ of binary variables $p_i$, $q_i \in \{0,1\}$ denoting the action of the projectors, we can express the final state as a sum of all possible trajectories as follows
\begin{equation}\label{eq:expn}
\begin{aligned}
    \E_U&\left[\rho_{k}^{sys} (\rho_{in})\right ]
     \equiv \sum_{\textbf{p},\textbf{q}\in\{0,1\}^{\otimes k}} \E_U [U^{-1} P_{p_k} U_k \ldots P_{p_1} U_1 \rho_s U_1^\dagger 
     \\& P_{q_1}\ldots U_k^\dagger P_{q_k} U^{-1 \dagger}]  \times \tr{\mathcal{H}(\textbf{p}) \,\rho_{env}\, \mathcal{H}(\textbf{q}) }.
\end{aligned}
\end{equation}
where evolution along a trajectory is denoted as $\mathcal{H}(\textbf{x}) = e^{-iH_{x_k}t}\ldots e^{-iH_{x_1}t}$.
As explained in the Supplementary Information, we can compute the average over the unitaries using methods of Haar integration and the \textit{average sequence fidelity}, the output of an RB experiment, in this case is shown to be
\begin{equation}
    \frac{1}{2}+\frac{1}{2}\,\frac{1}{6^k}\left( \sum_{\textbf{p},\textbf{q}} 2^{d(\textbf{p},\textbf{q})} \tr{\mathcal{H}(\textbf{p}) \,\rho_{env}\, \mathcal{H}^\dagger (\textbf{q}) }\right),
\end{equation}
where $d(\textbf{p},\textbf{q})$ is the Hamming distance between $\textbf{p}$ and $\textbf{q}$, and $\sum_{\textbf{p},\textbf{q}} 2^{d(\textbf{p},\textbf{q})} =6^k$.\\

\paragraph{Numerical Implementation}
In order to avoid the exponentially growing number of trajectories in the analytical expression in Eq.~\ref{eq:expn}, we developed a numerical method that is far more efficient in time. 
With every additional layer there is an additional unitary twirling and due to the nature of the interaction term in the Hamiltonian, this includes a step of projecting onto the $4^n$ computational basis states of $\mathcal{B}(\mathcal{H})$. After averaging we are left with $4^{2n}$ terms (see Supplementary Information), and the process continues. Knowing the state after every step explicitly in numerics allows us to circumvent this by regrouping the $4^{2n}$ terms into $4^n$ basis states and thereby always limiting the number of terms to sum. 
We would like to note that we imposed an occupational number cutoff for the environment for all our numeric implementations.
\\

\paragraph{Markovian Evolution}\label{subsec:Markovian}
With a Markovian assumption, as shown in Fig.~\ref{fig:circuit_model}(a), we assume the environment resets before every interaction with the system, thus carrying no information about the system with itself and preventing non-Markovian correlations.
We showed that average sequence fidelity in this case is an exponential decay with depth, which agrees with the existing literature. It is explicitly given by
\begin{equation}
     \frac{1}{2} + \frac{1}{2} \left(\frac{1+2\,\tr{cos(2gxt)\rho_{env}}}{3}\right)^k.
\end{equation}
\\

\paragraph{Dependence on gateset}
We analyze the fidelity decay for a model by considering a different gateset than the previous examples. We consider a single-qubit system initialized in the $|+\rangle$ state
and the unitary gates uniformly sampled from $\{X,\I\}$. Both $X$ and $\I$ gates leave a $|+\rangle$ state unchanged, eliminating the need for an inverse. Using the anti-commutation relation $\{X,Z\}=0$, the action of $X$ gate on the evolution operator is 
\begin{equation}
    Xe^{-iHt}=X\,e^{-i(g(a+a^\dagger)Z+\omega aa^\dagger\I)t}=e^{-i(-g(a+a^\dagger)Z+\omega aa^\dagger\I)t}\,X.
\end{equation}

Using this we showed that the average fidelity is
\begin{equation}
\begin{aligned}
    \E_{[k]} \mathrm{tr}_E &\left(\langle +| e^{-iH_kt} (|+\rangle\langle +| \otimes \rho_{env}) e^{iH_kt}|+\rangle\right) 
    \\& = \frac{1}{2} + \frac{1}{2} \tr{cos(2gxt)^k \rho_{env}}.
\end{aligned}
\end{equation}

\begin{figure}[tb]
    \includegraphics[width = .49 \textwidth]{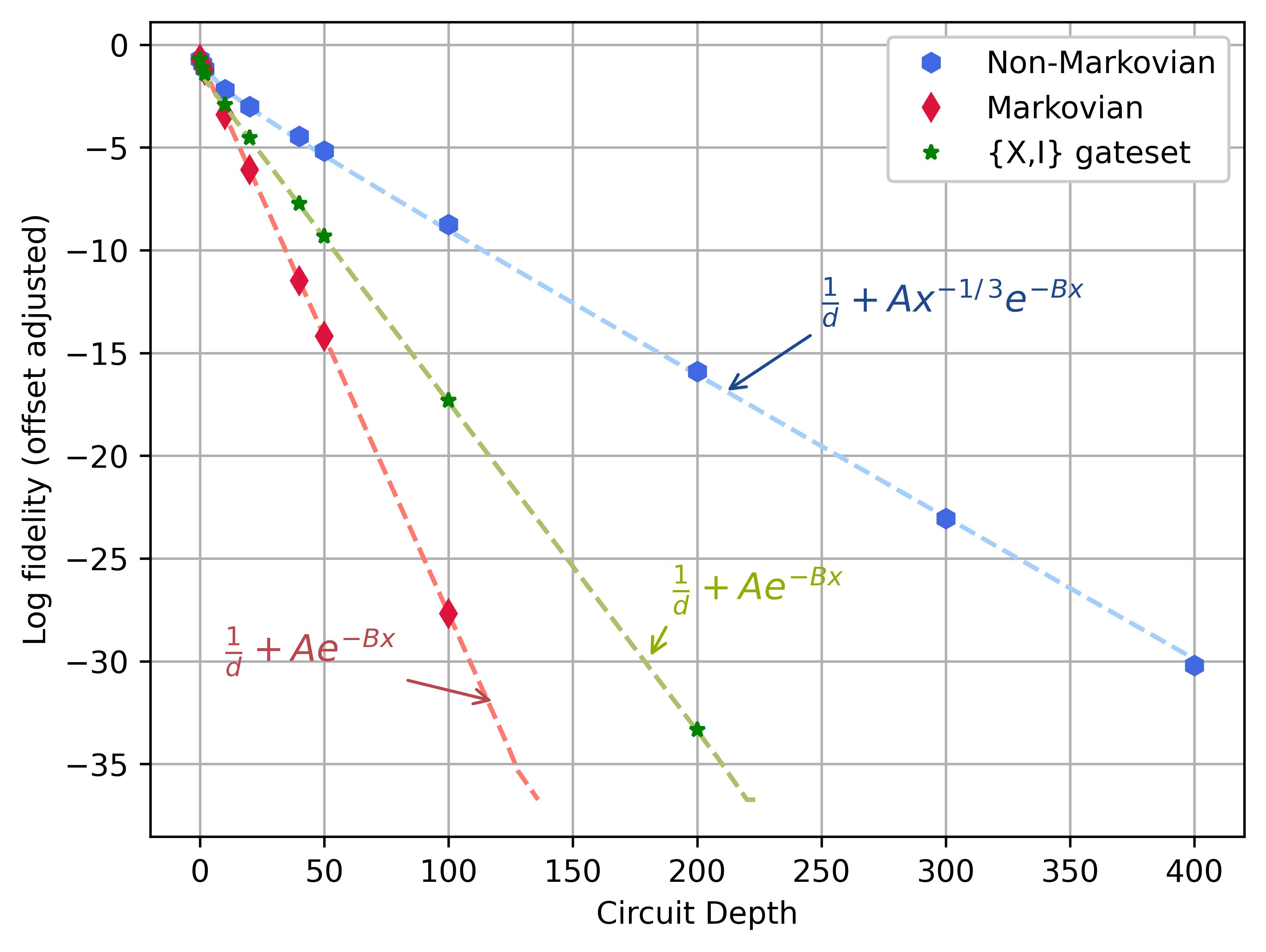}
    \caption{Single qubit decay of randomized benchmarking output with circuit depth , along with labelled numerical fits (on a log scale after removing the offset). $g$, $\omega$, temperature $ = 4,10,10^{-10}$ (relative units), time step= $0.1$s}
    \label{fig:rb_decay}
\end{figure}

The above three cases are illustrated in Fig.~\ref{fig:rb_decay}. Crucially, it leads us to the observation that RB decay of non-Markovian systems does not follow an exponential trend, and can be explained as a power law times exponential, w.r.t. depth. The power law behavior is especially apparent at smaller depths and provides us with a powerful signature of temporal correlations-- one that can be used to identify and characterize such noise. In the figure, non-Markovian decay is compared with the Markovian and the XI gateset cases, both of which exhibit exponential RB decays as proved.

\paragraph{Behavior of environment modes}
To glance into the change in the environment throughout this process, we look at how the average photon-number of environment, which begins in a thermal state, changes with the depth of the circuit. Starting with Eq.~\ref{eq:general_state}, and an initial state $\rho_{in}=\rho_s \otimes \rho_{env}$, the average photon-number after $k$ layers is
\begin{equation}\label{eq:avg_photon}
    \begin{aligned}
        \langle n_k \rangle &= \E[n_{k}^{env}(\rho_{in})] 
        \\ &= \sum_{\textbf{p},\textbf{q}\in\{0,1\}^{\otimes k}} \tr{\E_U[\M_k]} \,\tr{a^\dagger a \,\mathcal{H}(\textbf{p}) \rho_{env} \mathcal{H}^\dagger(\textbf{q})}
        \\ &= \sum_{\textbf{p}\in\{0,1\}^{\otimes k}} \frac{1}{2^k} \tr{\rho_s}\, \tr{a^\dagger a \,\mathcal{H}(\textbf{p}) \rho_{env} \mathcal{H}^\dagger(\textbf{p})}.
    \end{aligned}
\end{equation}

\begin{figure}[tb]
    \includegraphics[width = .49 \textwidth]{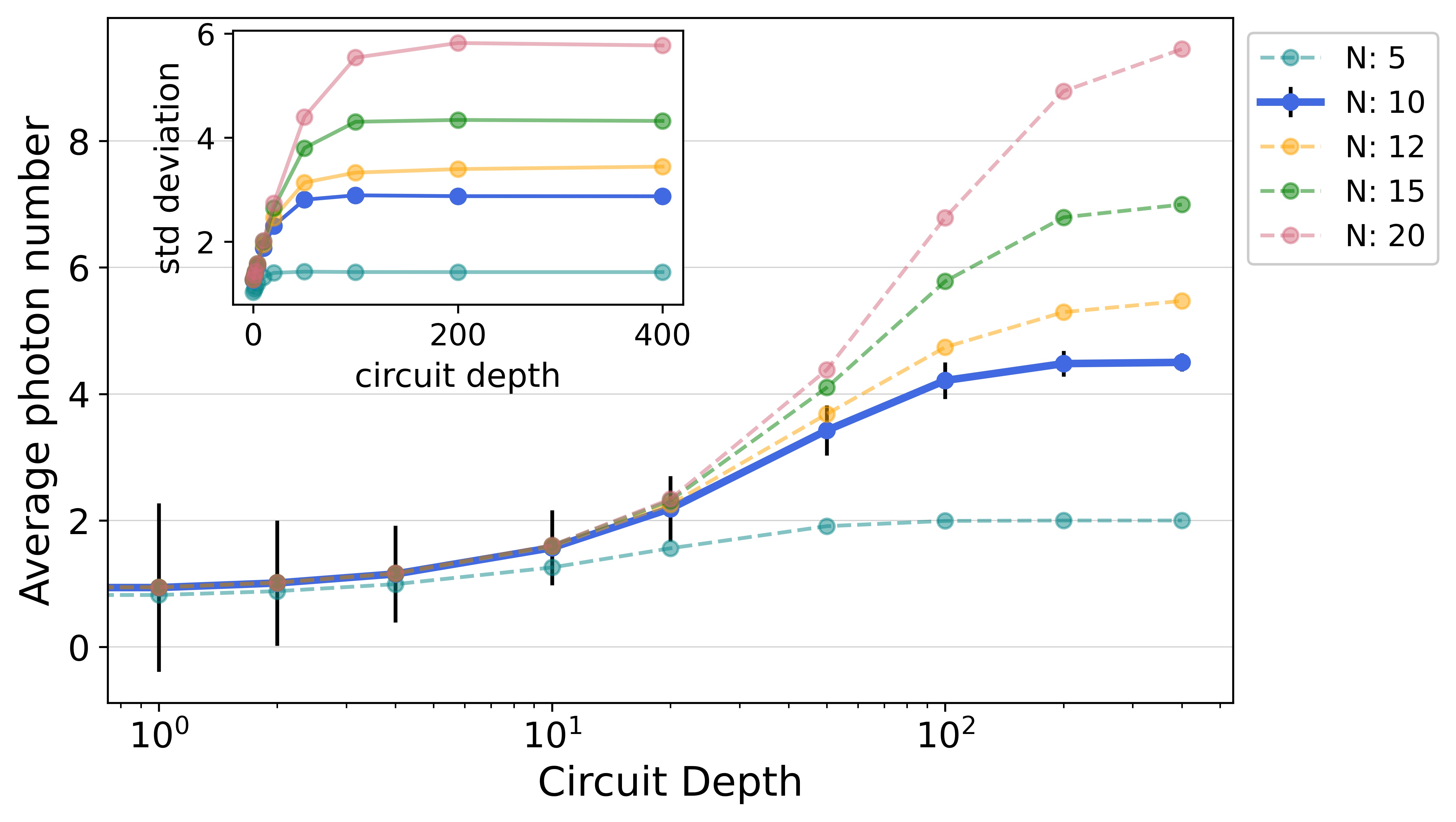}
    \caption{Average photon number of the environment initialized to a thermal state, starting at $\langle n\rangle \approx 0.87$, for different values of occupational-number cutoff for the environment mode. It saturates to a value of around $4.5$ for a cutoff value of $N=10$, see text for discussion. $g$, $\omega$, temperature $ = 4,10,10^{-10}$ (relative units), time step= $0.1$s for all the curves. \textit{Inset}- Variance of the photon number at different values of $N$.}
    \label{fig:avg_photon_num}
\end{figure}

Eq.~\ref{eq:avg_photon}, we observed that $\langle n \rangle$ saturates to a value independent of the parameters $g,\omega,t$. Fig.~\ref{fig:avg_photon_num} shows $\langle n \rangle$ has a strong dependence on the occupational number cutoff of the environment, indicating that the system is undergoing continuous heating in this model.  The slow overall rate of the heating compared to a Markovian system qualitatively explains the slower decay of the RB output in the non-Markovian case.

\paragraph{Signatures of non-Markovianity}
Several methods have been introduced to reliably identify and quantify non-Markovianity in quantum systems \cite{Rivas2014, Breuer2009, Hosseiny2024}; information backflow from environment to system, trace distance \cite{Laine2010}, divisibility of the evolution map \cite{Chruciski2011}, etc. However, these methods are considered independently from RB experiments.  Here, we introduce a readily measurable quantity in RB that can be used to determine the presence of non-Markovian noise in the system.

\begin{figure}[tb]
    \includegraphics[width = .49\textwidth]{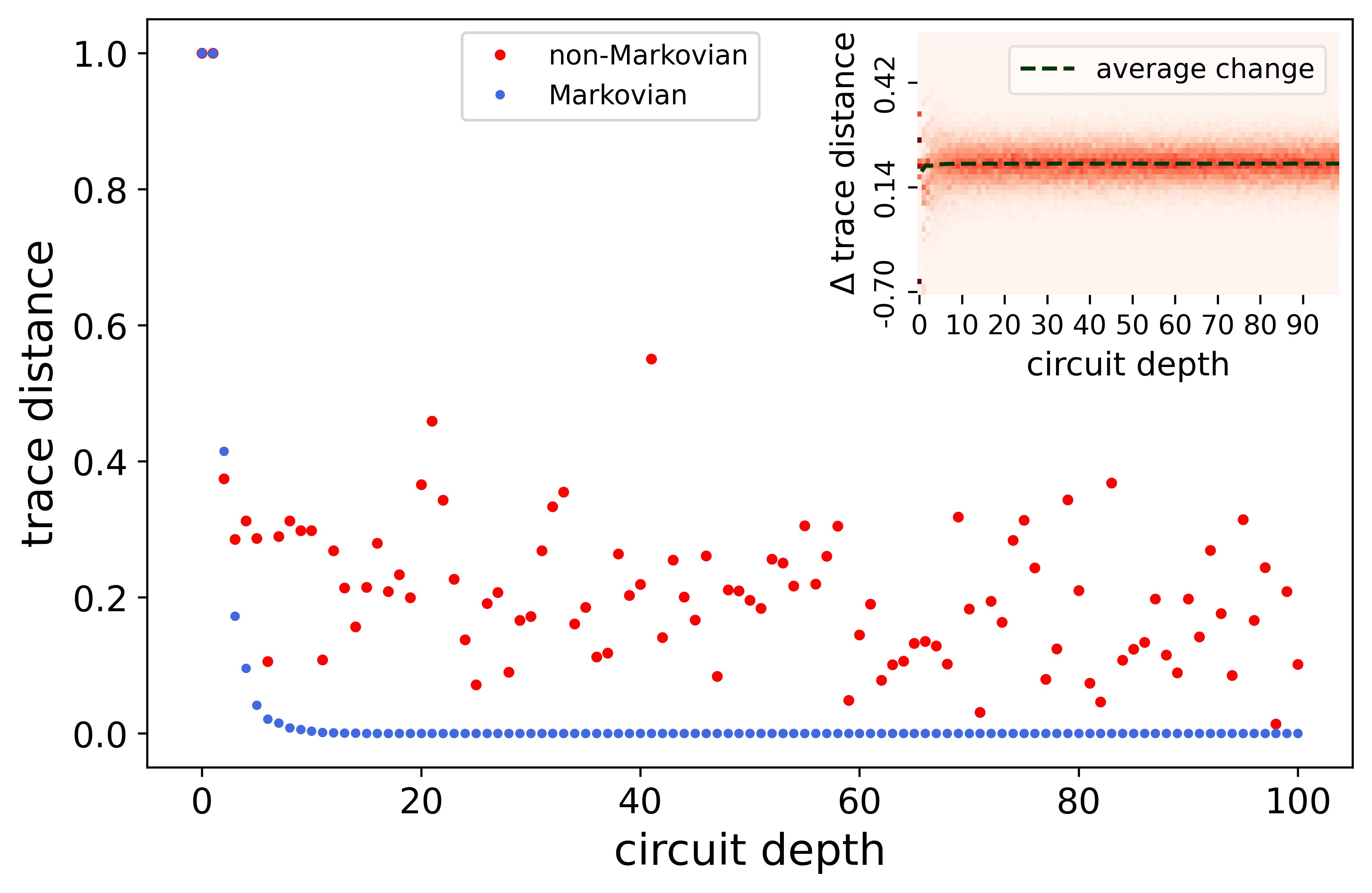}
    \caption{Difference in trace distance between the evolutions beginning in the orthogonal states $|0\rangle$ and $|1\rangle$, plotted wrt depth for a random circuit, for both Markovian and non-Markovian models. $g$, $\omega$, temperature $ = 4,10,10^{-10}$ (relative units), time step= $0.1$s. 
    \textit{Inset}- Distribution of the change in trace distance for non-Markovian circuits. The histograms, collected for $10^3$ random Clifford circuits, are displayed as color plots.
    }
    \label{fig:trace_dist}
\end{figure}

In particular, we propose to measure the trace distance between the evolution of orthogonal states ($|0\rangle$ and $|1\rangle$) at varying sequence lengths; a measure that must decrease monotonically for Markovian systems as both states depolarize to the same maximally mixed state. The evolution map here is one instance of the the non-Markovian model in Fig.~\ref{fig:circuit_model}, with random unitary gates.
\begin{equation}
    D(\rho^{sys,0}_k,\rho^{sys,1}_k) = \frac{1}{2} \tr{|\rho^{sys,0}_k-\rho^{sys,1}_k|}
\end{equation}
For each random circuit instance, we observed distinct time steps displaying an increase in trace distance, see Fig.~\ref{fig:trace_dist}, a clear indication of the presence of non-Markovianity. In contrast, the trace distance is monotonically decreasing for a Markovian circuit. The inset in Fig.~\ref{fig:trace_dist} shows the distribution of the change in trace distance over many circuit realizations.

See Supplementary Information for a discussion on the mixed state fidelity between the two evolutions.

\paragraph{Multi-qubit systems and multimode baths}
Our results extend to, and result in similar trends, in multi-qubit systems and systems interacting with multi-mode baths. One can write a general Hamiltonian for $n$ qubits and $m$ modes as follows

\begin{equation}
    \begin{aligned}
        H &= 
        \sum_i  \omega_i a^\dagger_i a_i + \sum_{ij} g_{ij} \sigma_{zj}(a_i+a^\dagger_i)  = \bigoplus_{p\in \{0,1\}^{\otimes n}}  H_{p} ,
    \end{aligned}
\end{equation}
where, using the weighted sum of a bit string $w(q)\equiv \sum_{j=1}^n (-1)^{q_j}$,
\begin{equation}
    H_{p} = \sum_{i=1}^m \, \omega_i a^\dagger_i a_i + w(p)\, g_{ij}  (a_i+a^\dagger_i).
\end{equation}

As before we can assume that we begin in a product state $\rho_{SE} = \rho_s \otimes \rho_{env}$, and apply a sequence of $n$-qubit unitary operations $U_1,\ldots,U_k,U_{inv}$, where $d=2^n$, $U_i\in \mathbb{U}_d$, and $U_{inv}=(U_1\ldots U_k)^{-1}$.  We show in the Supplementary Information that the RB decays to $\frac{1}{d}$.

\paragraph{Conclusions}
We studied a system interacting with an environment with a long quantum memory, resulting in a non-Markovian dynamics of the system. We found a drastically different RB decay with non-Markovian noise compared to the Markovian case, that differ qualitatively at early depths. Although they both tail off exponentially, non-Markovian decays in our model are much slower than ones for Markovian systems, under the same parameters. We showed this by considering a simplified, exactly solvable model of qudits interacting with a Bosonic bath. By determining the signatures of a non-Markovian decay, we provide guidance to identify these effects in common characterization experiments.
The computation in this work was possible because the eigenvectors of the Hamiltonian could be solved analytically. In general, this simplification is not  possible, making it important to determine the universality of these results to other models of non-Markovian quantum systems.
This work opens interesting avenues for research to theoretically relate the source of the non-exponential behavior to the details of the system-bath interaction, paving the way for more precise noise characterization.
\\

\begin{acknowledgments}
   We thank Xiao Xiao and Anantha Rao for useful discussions.  This work was supported in part by ARO grant W911NF-23-1-0258.
\end{acknowledgments}

\bibliographystyle{apsrev4-2}
\bibliography{main.bib}

\end{document}


\title{Supplementary Information for ``Quantum non-Markovian noise in randomized benchmarking of spin-boson models"}
\date{\today}
\onehalfspacing

\maketitle

\subsection{Non-Markovian Evolution}
Our objective is to find the final state of the system, after $k$ time steps, averaged over the unitary group $\mathcal{U}(d)$. Each Hamiltonian interaction has a duration $t$ (see Fig.~1 in main text). 
\begin{equation}\label{eq:general_state}
\begin{aligned}
    \E_U\left[\rho_{k}^{sys}\right] = tr_E ( \E_U[ U^{-1} e^{-iHt}U_k \ldots e^{-iHt}U_1\, \rho_s \otimes \rho_{env} U_1^\dagger e^{iHt}\ldots U_k^\dagger e^{iHt} U^{-1 \dagger}])
\end{aligned}
\end{equation}

Note that, by linearity, if the initial state is a mixed state  $\rho_{in} = \sum_a c_a\, \rho_{s,a} \otimes \rho_{env,a}$, the final state can we determined by averaging each product term separately
\begin{equation}\label{eq:mixed_st}
    \E_U[\rho_{f,k} (\rho_{in})] = \sum_a c_a \, \E_U[\rho_{f,k}(\rho_{s,a}\otimes \rho_{env,a})].
\end{equation}
Therefore we assume the initial state is a product state with the results extending to mixed states. We compute the average over unitaries using the expression for Haar average shown below. By definition this is also valid for any 2-design, and therefore the Clifford group which is of interest in RB. Using
\begin{equation}\label{eq:haar}
\begin{aligned}
\int_{\mathcal{U}(d)} &d \eta(U) U^{\dagger} A U \rho U^{\dagger} B U=\frac{\tr{A B} \tr{\rho}}{d} \frac{\I}{d}
+\left(\frac{d \tr{A} \tr{b}-\tr{A B}}{d\left(d^2-1\right)}\right)\left(\rho-\tr{\rho} \frac{\I}{d}\right)~,
\end{aligned}
\end{equation}
and the projectors $P_0 = |0\rangle\langle 0|$ and $P_1 = |1\rangle\langle 1|$, a single layer of twirling can be simplified as  
\begin{equation}\label{eq:one_layer}
\begin{aligned}
    \E_U [U^{-1} P_p U \rho_s U^\dagger P_q U^{-1 \dagger} ] 
    &= \E_U [U^\dagger P_p U \rho_s U^\dagger P_q U] \\
    &= \frac{\tr{P_p P_q} \tr{\rho_s}}{2} \frac{\I}{2} + \left(\frac{2-\tr{P_p P_q}}{6}\right)\left(\rho_s - \tr{\rho_s} \frac{\I}{2}\right) \\
    &= \left( \frac{2}{3}\delta_{pq} -\frac{1}{3}\right) \tr{\rho_s} \,\frac{\I}{2} + \left( \frac{2-\delta_{pq}}{6}\right)\rho_s.
\end{aligned}
\end{equation}

For $k$ layers, let us define $\M_k(\textbf{p},\textbf{q})$ as the state of system determined by projector sequences $\textbf{p}=(p_1,\ldots,p_k)$ and $\textbf{q}=(q_1,\ldots,q_k)$     
\begin{widetext}

\begin{equation}\label{eq:m_expn}
\begin{aligned}
    \E_U [\M_k(\textbf{p},\textbf{q})] &\equiv \E_U [U^{-1} P_{p_k} U_k \ldots P_{p_1} U_1 \rho_s U_1^\dagger P_{q_1}\ldots U_k^\dagger P_{q_k} U^{-1 \dagger}] 
    \\ &= \E_U [(U_k \ldots U_1)^\dagger P_{p_k}(U_k \ldots U_1) \ldots (U_2 U_1)^\dagger P_{p_2}(U_2 U_1) U_1^\dagger P_{p_1}U_1\rho_s 
    \\ & \qquad \times U_1^\dagger P_{q_1} U_1 (U_2 U_1)^\dagger P_{q_2}\ldots (U_k\ldots U_1)^\dagger P_{q_k} (U_k\ldots U_1)]
    \\ &= \E_U [\Tilde{U}_k^\dagger P_{p_k} \Tilde{U}_k \ldots \Tilde{U}_1^\dagger P_{p_1}\Tilde{U}_1 \rho_s \Tilde{U}_1^\dagger P_{q_1} \Tilde{U}_1\ldots \Tilde{U}_k^\dagger P_{q_k} \Tilde{U}_k]
    \\& =\E_{[k-1]} \left[\E_{k}[\Tilde{U}_k^\dagger P_{p_k} \Tilde{U}_k \M_{k-1}\, \Tilde{U}_k^\dagger P_{q_k} \Tilde{U}_k]\right]
    \\& =\E_{[k-1]}\left[ \left( \frac{2}{3}\delta_{p_k q_k} -\frac{1}{3}\right) \tr{\M_{k-1}} \,\frac{\I}{2} + \left( \frac{2-\delta_{p_k q_k}}{6}\right)\M_{k-1} \right],
\end{aligned}
\end{equation}
\end{widetext}

 where $\M_i = \Tilde{U}^\dagger_{i} P_{p_{i}}\Tilde{U}_{i} \M_{i-1} \Tilde{U}^\dagger_{i}P_{q_{i}}\Tilde{U}_{i}$ is recursively defined and $\M_0 = \rho_s$. We used $U^{-1}=U^\dagger$, inserted unitaries and their inverses in the first step to separate and cast the averages of different unitaries into the form given by Eq.~\ref{eq:haar}. We denote $\prod_{i=0}^{m-1} U_{m-i} \equiv \Tilde{U}_m$ and the expectation $\E_{U_1,\ldots,U_{i}} \equiv \E_{[i]}$, making $\E_{[k]} = \E_U$. 
 
 Taking trace on both sides and denoting $\delta_{p_i,q_i}$ as $\delta_i$,
 \begin{equation}\label{eq:m_tr}
     \begin{aligned}
         \E_{[k]}\left[\tr{\M_k}\right] &= \E_{[k-1]}\left[ \left( \frac{2}{3}\delta_{p_k q_k} -\frac{1}{3}\right) \tr{\M_{k-1}} + \left( \frac{2-\delta_{p_k q_k}}{6}\right) \tr{\M_{k-1}} \right]
         \\ &= \frac{\delta_k}{2} \, \E_{[k-1]}\left[\tr{\M_{k-1}}\right]
         \\ &= \prod_{j=1}^k \frac{\delta_j}{2} \, \tr{\M_0} 
         \\ &= \frac{1}{2^k} ~\left(\prod_{j=1}^k\delta_j\right) ~ \tr{\rho_s}.
     \end{aligned}
 \end{equation}

Substituting this as the first term of Eq.~\ref{eq:m_expn},
\begin{widetext}
\begin{equation}
    \begin{aligned}
        \E_{[k]}[\M_k] &=  \left( \frac{2}{3}\delta_k -\frac{1}{3}\right) \E_{[k-1]}[\tr{\M_{k-1}}] ~\tr{\rho_s}~\frac{\I}{2} + \left( \frac{2-\delta_{p_k q_k}}{6}\right) \E_{[k-1]}[\M_{k-1}]
        \\ &= \frac{1}{3}\left( 2\delta_k -1\right) \left(\prod_{j=1}^{k-1} \frac{\delta_j}{2}\right)\tr{\rho_s}\frac{\I}{2} + \frac{1}{3}\left( 1-\frac{\delta_k}{2}\right) \E_{[k-1]}[\M_{k-1}].
    \end{aligned}
\end{equation}
\end{widetext}

Applying it $k$ times recursively,
\begin{equation}
    \begin{aligned}
        \E_{[k]}[\M_k] &=  ~\Delta(\textbf{p},\textbf{q}) ~\tr{\rho_s} ~\frac{\I}{2} + \frac{1}{3^k} \prod_{i=1}^k \left(1-\frac{\delta_i}{2}\right) ~\rho_s
        \\ &\equiv  \left(\Delta(\textbf{p},\textbf{q})~\tr{\rho_s} ~\frac{\I}{2} + \frac{1}{3^k} \gamma\,(\textbf{p},\textbf{q})~\rho_s \right),
    \end{aligned}
\end{equation}
where $\Delta(\textbf{p},\textbf{q})$ can be determined by taking trace and using Eq.~\ref{eq:m_tr},
\begin{equation}
    \frac{1}{2^k} \left(\prod_{j=1}^k\delta_j\right) \, \tr{\rho_s} = ~\Delta(\textbf{p},\textbf{q}) ~\tr{\rho_s} + \frac{1}{3^k} \gamma\,(\textbf{p},\textbf{q}) ~\tr{\rho_s},
\end{equation}
which gives us
\begin{equation}
    \Delta(\textbf{p},\textbf{q}) = \frac{1}{2^k}\prod_{j=1}^k\delta_j - \frac{1}{3^k}\gamma\,(\textbf{p},\textbf{q}).
\end{equation}
Writing $\prod_{j=1}^k\delta_j$ as $\delta_{\textbf{p},\textbf{q}}$, we have
\begin{equation}\label{eq:rho_system}
\begin{aligned}
    \E_U[\M_k(\textbf{p},\textbf{q})] = \frac{1}{3^k} &\gamma\,(\textbf{p},\textbf{q}) \left(\rho_s - \tr{\rho_s}\frac{\I}{2}\right) 
    \\& + \frac{1}{2^k} ~\delta_{\textbf{p},\textbf{q}} ~\tr{\rho_s}\frac{\I}{2}.
\end{aligned}
\end{equation}

\subsection{Markovian Evolution}
When the state of the environment is refreshed before every interaction with the system, the resulting system state after $k$ time-steps is
\begin{widetext}
\begin{equation}\label{eq:mark}
\begin{aligned}
    \E_U\left[\rho_{k}^{sys, M}\right ] &= \sum_{\textbf{p},\textbf{q}\in\{0,1\}^{\otimes k}} \E_U [U^{-1} P_{p_k} U_k \ldots P_{p_1} U_1 \rho_s U_1^\dagger P_{q_1}\ldots U_k^\dagger P_{q_k} U^{-1 \dagger}] 
    \\ & \hspace{12em} \times \prod_{j=1}^k \tr{e^{-iH_{p_j}t} \rho_{env} e^{iH_{q_j}t} }
    \\& \text{Using Eq.~\ref{eq:rho_system},}
    \\& = \sum_{\textbf{p},\textbf{q}\in\{0,1\}^{\otimes k}} \frac{1}{3^k} \gamma(\textbf{p},\textbf{q}) \left(\rho_s - \tr{\rho_s}\frac{\I}{2}\right)\prod_{j=1}^k \tr{e^{-iH_{p_j}t} \rho_{env} e^{iH_{q_j}t}}  
    \\&\hspace{12em}+ \sum_{\textbf{p}}\frac{1}{2^k} ~\tr{\rho_s}~\tr{\rho_{env}}^k \frac{\I}{2}
    \\& =\frac{1}{3^k}\left(\rho_s - \tr{\rho_s}\frac{\I}{2}\right) \sum_{\textbf{p},\textbf{q}\in\{0,1\}^{\otimes k}} \prod_{j=1}^k \left(1-\frac{\delta_j}{2}\right) \tr{e^{-iH_{p_j}t} \rho_{env} e^{iH_{q_j}t} } 
    \\&\hspace{12em} +~\tr{\rho_s}~\tr{\rho_{env}}^k \frac{\I}{2} 
    \\& =\frac{1}{3^k}\left(\rho_s - \tr{\rho_s}\frac{\I}{2}\right) \prod_{j=1}^k \sum_{p_i,q_i\in\{0,1\}} \left(1-\frac{\delta_j}{2}\right) \tr{e^{-iH_{p_j}t} \rho_{env} e^{iH_{q_j}t} }  
    \\&\hspace{12em} +~\tr{\rho_s}~\tr{\rho_{env}}^k \frac{\I}{2}  
    \\ &=\frac{1}{3^k} \left(\tr{\rho_{env}} + 2\,\tr{cos((H_0-H_1)t)\,\rho_{env}}\right)^k \left(\rho_s - \tr{\rho_s}\frac{\I}{2}\right) + \tr{\rho_s}~\tr{\rho_{env}}^k \frac{\I}{2}   
\end{aligned}
\end{equation}
\end{widetext}
\noindent
When $\tr{\rho_{env}}=1$, the prefactor $\frac{1}{3^k}\, (1+2\,\tr{cos((H_0-H_1)t) \rho_{env}})^k$ exponentially decays with depth and therefore the final state approaches the maximally mixed state $\frac{\I}{2}$.

Using $\tr{\rho_{env}}=1$, $\tr{\rho_s}=1$, and $H_0-H_1 = 2gx$, RB output is
\begin{equation}
    \tr{\E_U\left[\rho_{k}^{sys, M}\right ] \, \rho_s} = \frac{1}{2} + \frac{1}{2} \left(\frac{1+2\,\tr{cos(2gxt)\rho_{env}}}{3}\right)^k
\end{equation}

\subsection{XI Model}
For a sequence of length $k$, let the variables $\{u_i\}_{i=1}^k$ represent the gate at the $i^{th}$ position: with $0$ for $\I$ gate and $1$ for $X$ gate. Then, if the number of $X$ gates, i.e. $\sum_i u_i$, is equal to $N_X$, the final state can be written as 

\begin{equation}
\begin{aligned}
    U_k e^{-iHt}\ldots & U_1 e^{-iHt} (|+\rangle\langle +| \otimes \rho_{env}) U_1^\dagger e^{iHt}\ldots U_k^\dagger e^{iHt} 
    \\ & \approx e^{-i(-gf_k\,Z(a+a^\dagger)+ k\omega \,a^\dagger a)t} \, X^{N_X}\, (|+\rangle\langle +| \otimes \rho_{env}) 
    \\& \quad\quad X^{N_X}\, e^{i(-gf_k\,Z(a+a^\dagger)+ k\omega \,a^\dagger a)t}
    \\ &\equiv e^{-iH_kt} (|+\rangle\langle +| \otimes \rho_{env}) e^{iH_kt}.
\end{aligned}
\end{equation}
Note that the unitaries acting on the system $U_i\otimes \I_{env}$ are written as $U_i$ for simplicity. The phase factor introduced by moving all the $X$ operations to the right is given by $f_k = (-1)^{u_1} + (-1)^{u_1+u_2} +\ldots+ (-1)^{u_1+\ldots+u_k}$. Defining an effective Hamiltonian $H_k$ using $g_k\equiv gf_k$ and $\omega_k\equiv k\omega$, we can proceed as with the earlier cases by splitting it as a sum $H_k = H_{0,k} \oplus H_{1,k}$, where the terms act on $|0\rangle$ and $|1\rangle$ qubit states respectively. Using $|+\rangle\langle +| = \frac{1}{4} \,(|0\rangle\langle 0| + |0\rangle\langle 1| + |1\rangle\langle 0| + |1\rangle\langle 1|)$, the sequence fidelity is

\begin{equation}\label{eq:xi}
    \begin{aligned}
        \mathrm{tr}_E &\left(\langle +| e^{-iH_kt} (|+\rangle\langle +| \otimes \rho_{env}) e^{iH_kt}|+\rangle\right) \\ &= \frac{1}{4} \mathrm{tr}_E\left(e^{-iH_{k,0}t}\rho_{env}e^{iH_{k,0}t} + \ldots + e^{-iH_{k,1}t}\rho_{env}e^{iH_{k,1}t} \right)
        \\ &= \frac{1}{4} \left(2+ \tr{e^{iH_{k,1}t}e^{-iH_{k,0}t} \rho_{env}} + \tr{e^{iH_{k,0}t}e^{-iH_{k,1}t} \rho_{env}} \right)
        \\ &= \frac{1}{2} \left(1+ \tr{cos((H_{k,0}-H_{k,1})t) \rho_{env}}  \right)
        \\ &= Re\{\,\frac{1}{2} \left(1+ \tr{e^{-i(H_{k,0}-H_{k,1})t} \rho_{env}}  \right)\}
        \\ &= Re\{\,\frac{1}{2} \left(1+ \tr{e^{-2igf_k xt} \rho_{env}}  \right)\}.
    \end{aligned}
\end{equation}

Note that $f_k = (-1)^{u_1} + (-1)^{u_1+u_2} +\ldots+ (-1)^{u_1+\ldots+u_k} = (-1)^{u_1} \, (1+f_{k-1})$, and so on. To determine the average output, we can therefore sequentially average from $u_1$ through $u_k$, giving us
\begin{equation}\label{eq:avg_phase}
\begin{aligned}
      \E_{[k]} e^{-2igf_k xt} &= \frac{1}{2} \E_{[k-1]} \left(e^{-2ig(1+f_{k-1})xt} +  e^{2ig(1+f_{k-1})xt} \right)
      \\ &= Re\{\, \E_{[k-1]} e^{-2ig(1+f_{k-1}) xt} \}
      \\ &= Re\{\, e^{-2igxt} \,\E_{[k-1]} e^{-2igf_{k-1} xt}\}
      \\ &= Re\{\, e^{-2igxt}\}^k
      \\ &= cos(2gxt)^k.
\end{aligned}
\end{equation}
Using Eq.~\ref{eq:xi} and Eq.~\ref{eq:avg_phase}, the average sequence fidelity for the XI gateset is 
\begin{equation}
    \E_{[k]} \mathrm{tr}_E \left(\langle +| e^{-iH_kt} (|+\rangle\langle +| \otimes \rho_{env}) e^{iH_kt}|+\rangle\right) 
     = \frac{1}{2} + \frac{1}{2} \tr{cos(2gxt)^k \rho_{env}}.
\end{equation}

\subsection{Multi-qubit systems and multimode baths}
Consider an $n$-qubit system interacting with a multimode Bosonic environment. Let us start with a product state $\rho_{SE} = \rho_s \otimes \rho_{env}$, and apply a sequence of $n$-qubit unitary operations $U_1,\ldots,U_k,U_{inv}$, where the dimension of the system $d=2^n$, $U_i\in \mathbb{U}_d$, and $U_{inv}=(U_1\ldots U_k)^{-1}$. We want to compute the average sequence fidelity 

\begin{equation}
\begin{aligned}
    \sum_{\substack{p_i,q_i \in \{0,1\}^{\otimes n} \\ 1\leq i \leq k}} \E_U &[\Tilde{U}_k^\dagger P_{p_k} \Tilde{U}_k \ldots \Tilde{U}_1^\dagger P_{p_1}\Tilde{U}_1 \rho_s \Tilde{U}_1^\dagger P_{q_1} \Tilde{U}_1
    \\& \ldots \Tilde{U}_k^\dagger P_{q_k} \Tilde{U}_k] \otimes \, \mathcal{H}(\textbf{p}) \rho_{env} \mathcal{H}^\dagger(\textbf{q}),
\end{aligned}
\end{equation}
where $\mathcal{H}(\textbf{p})=e^{-iH_{\textbf{p}}t}$, and as defined in the main text $H_{p} = \sum_{i=1}^m \, \omega_i a^\dagger_i a_i + w(p)\, g_{ij}  (a_i+a^\dagger_i)$, using the weighted sum of a bit string $w(q)\equiv \sum_{j=1}^n (-1)^{q_j}$.


Similar to the single-qubit analysis, Haar average at $d=2^n$ is
\begin{equation}
    \begin{aligned}
        \E_U [U^\dagger P_p U \M U^\dagger P_q U] 
         &= \frac{\delta_{pq}}{d} \tr{\M} \frac{\I}{d} + \frac{1}{d^2 -1}(1-\frac{\delta_{pq}}{d}) \left( \M-\tr{\M}\frac{\I}{d} \right)
        \\ &= \frac{d \delta_{pq} -1}{d^2 -1} \,\tr{\M} \frac{\I}{d} + \frac{1}{d^2 -1}\,(1-\frac{\delta_{pq}}{d}) \M.
    \end{aligned}
\end{equation}

\noindent Defining $\M_k (\textbf{p},\textbf{q})$ as before, 
\begin{equation}
    \E_U[\tr{\M_k (\textbf{p},\textbf{q})}]= \prod_{j=1}^k \frac{\delta_j}{d} \,\,\tr{\rho_s},
\end{equation}

\begin{equation}
    \E_U[\M_k (\textbf{p},\textbf{q})] = \Delta(\textbf{p},\textbf{q}) \, \tr{\rho_s} \frac{\I}{d} + \frac{1}{(d^2-1)^k} \prod_{j=1}^k \left(1-\frac{\delta_j}{d}\right) \rho_s, \quad\text{and}
\end{equation}

\begin{equation}
    \Delta(\textbf{p},\textbf{q}) = \frac{1}{d^k} \prod_{j=1}^k \delta_j - \frac{1}{(d^2-1)^k} \prod_{j=1}^k \left(1-\frac{\delta_j}{d}\right).
\end{equation}

If $\tr{\rho_s}=\tr{\rho_{env}}=1$, and if the system begins in the ground state, RB output is 
\begin{equation}
    \frac{1}{d} +  \frac{1}{(d^2-1)^k}\, \sum_{\textbf{p},\textbf{q}}  \left(1-\frac{1}{d}\right)^{1+k-d(\textbf{p},\textbf{q})} \, \tr{\mathcal{H}(\textbf{p}) \rho_{env} \mathcal{H}^\dagger(\textbf{q})}.
\end{equation}


\begin{figure}[t]
    \includegraphics[width=\textwidth]{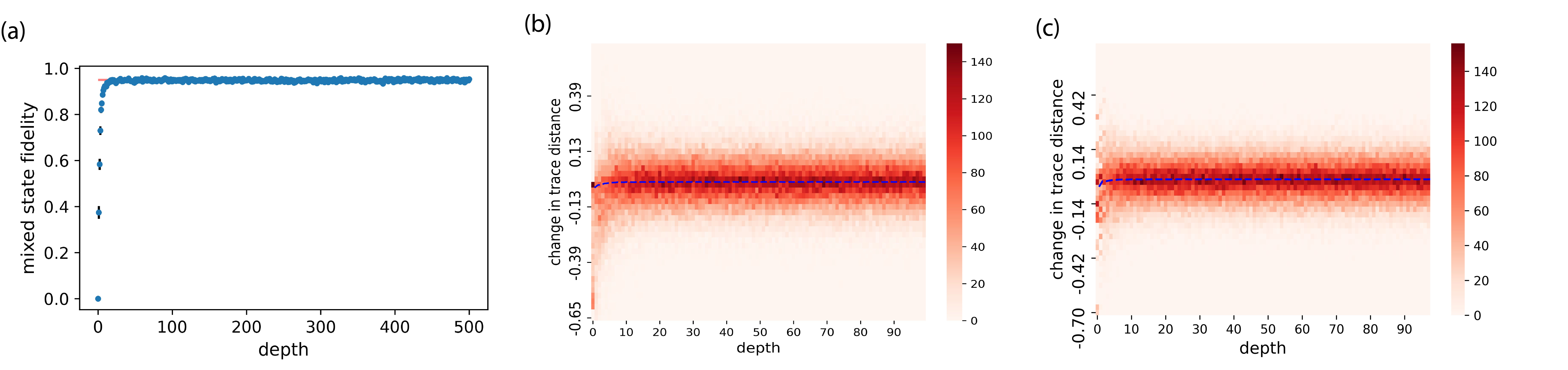}
    \caption{(a) Mixed state fidelity between the final states with $|0\rangle$ and $|1\rangle$ initial states. (b,c) Distribution of the change in trace distance, when the unitary gates are sampled from the Haar distribution and the Clifford group respectively. The histograms, collected for $10^3$ random circuits, are displayed as color plots. }
    \label{fig:mixed_state_fidelity}
\end{figure}

\subsection{Mixed state fidelity}
We employed trace distance to study the difference between the evolutions beginning in the orthogonal states $|0\rangle$ and $|1\rangle$ and to provide indications of non-Markovianity. Here, we look at the mixed state fidelity between the same two evolutions. Mixed state fidelity between two states $\rho$ and $\sigma$ is defined as
\begin{equation}
    F(\rho,\sigma) = \tr{\sqrt{\sqrt{\rho} \sigma \sqrt{\rho}}}^2,
\end{equation}
and can range from $0$ to $1$, as the states go from identical to orthogonal. In Fig.~\ref{fig:mixed_state_fidelity}(a) we can see that mixed state fidelity between these two evolutions saturates to a value close to 1, never reaching the maximum value. We note that this is because the final states closely fluctuate around the maximally mixed state without converging.

\bibliographystyle{apsrev4-2}
\bibliography{main.bib}